\documentclass[aps,prl,superscriptaddress,amsfonts,amsmath,amssymb,reprint,showpacs,floatfix]{revtex4-1}

\usepackage{url}
\usepackage{bm}
\usepackage{graphicx}
\usepackage{amsmath}
\usepackage{amstext}
\usepackage{amssymb}
\usepackage{amsfonts}
\usepackage{amsbsy}
\usepackage{verbatim}
\usepackage{color}
\usepackage[colorlinks=true, urlcolor=blue, linkcolor=blue, citecolor=blue, pdftex]{hyperref}
\usepackage{multirow}
\usepackage{pdfpages}
\usepackage{floatrow}

\begin{document}

\title{Vanishing spin gap in a competing spin-liquid phase in the kagome Heisenberg antiferromagnet}

\author{\href{http://www.ictp.it/research/cmsp/members/postdoctoral-fellows/yasir-iqbal.aspx}{Yasir Iqbal}}
\email[]{yiqbal@ictp.it}
\affiliation{The Abdus Salam International Centre for Theoretical Physics, P.O.~Box 586, I-34151 Trieste, Italy}
\author{\href{http://www.lpt.ups-tlse.fr/spip.php?article32}{Didier Poilblanc}}
\email[]{didier.poilblanc@irsamc.ups-tlse.fr}
\affiliation{Laboratoire de Physique Th\'eorique UMR-5152, CNRS and Universit\'e de Toulouse, F-31062 Toulouse, France}
\author{\href{http://people.sissa.it/~becca/}{Federico Becca}}
\email[]{becca@sissa.it}
\affiliation{Democritos National Simulation Center, Istituto Officina dei Materiali del CNR and 
SISSA-International School for Advanced Studies, Via Bonomea 265, I-34136 Trieste, Italy}

\date{\today}

\begin{abstract}
We provide strong numerical evidence, using improved variational wave functions, for a ground state 
with vanishing spin gap in the spin-$1/2$ quantum Heisenberg model on the kagome lattice. Starting 
from the algebraic $U(1)$ Dirac spin liquid state proposed by Y. Ran {\it et al.} 
[\href{http://dx.doi.org/10.1103/PhysRevLett.98.117205}{Phys. Rev. Lett. {\bf 98}, 117205 (2007)}] 
and iteratively applying a few Lanczos steps, we compute the lowest $S=2$ excitation 
constructed by exciting spinons close to the Dirac nodes. Our results are compatible with a vanishing 
spin gap in the thermodynamic limit and in consonance with a power-law decay of long distance 
spin-spin correlations in real space. The competition with a gapped (topological) spin liquid is 
discussed.
\end{abstract}


\maketitle

{\it Introduction}.
It is traditional wisdom that at low temperatures phases of matter condense 
by spontaneously breaking some symmetry and thus developing order. The possibility 
of realizing phases which evade ordering has only been fruitfully explored in the 
recent past with many exotic scenarios. Among these, the spin-$1/2$ quantum Heisenberg
antiferromagnet on frustrated lattices occupies a distinguished position. 
The kagome lattice represents one of the most promising candidates, given its large
degeneracy of the classical ground-state manifold and the strong quantum fluctuations.
Several studies in the past highlighted the difficulty in reaching a definitive
understanding of its low-energy properties~\cite{Lecheminant-1997,Sindzingre-2009,Nakano-2011,Lauchli-2011,Singh-1992,Misguich-2005,Balents-2010}.
Indeed, the identification of the precise nature of the ground state of the kagome 
spin-$1/2$ Heisenberg model remains unsettled and widely debated. Different approximate numerical
and analytical techniques have claimed a variety of ground states. On the one hand, 
density-matrix renormalization group (DMRG), pseudofermion functional 
renormalization group, and Schwinger boson mean-field calculations have supported a fully 
gapped $\mathbb{Z}_{2}$ topological spin-liquid ground state that does not break any lattice 
symmetry~\cite{Yan-2011,Depenbrock-2012,Nishimoto-2013,Ju-2013,Suttner-2013,Sachdev-1992,Wang-2006,Messio-2012}. 
On the other hand, a gapless (algebraic) and fully symmetric $U(1)$ Dirac spin liquid has been 
proposed as the ground state and widely studied using variational Monte Carlo
approach~\cite{Hastings-2000,Ran-2007,Hermele-2008,Ma-2008,Iqbal-2011a,Iqbal-2011b,Iqbal-2012,Iqbal-2013}. 
In addition, valence bond crystals of different unit cell sizes and symmetries have been 
suggested from other techniques~\cite{Marston-1991,Zeng-1995,Syro-2002,Nikolic-2003,Singh-2007,Poilblanc-2010,Evenbly-2010,Huh-2011,Hwang-2011,Poilblanc-2011,Capponi-2013}. 
The coupled-cluster method suggested a $q=0$ (uniform) state~\cite{Gotze-2011}. Finally, extending 
the construction of tensor network {\it Ans\"atze} of gapped $\mathbb{Z}_2$ spin 
liquids~\cite{Poilblanc-2012}, a recent calculation, based upon the so-called projected
entangled simplex states (PESS)~\cite{Xie-2013} which preserve lattice 
symmetries, gave remarkably accurate energies.

\begin{figure}[b]
\includegraphics[width=1.0\columnwidth]{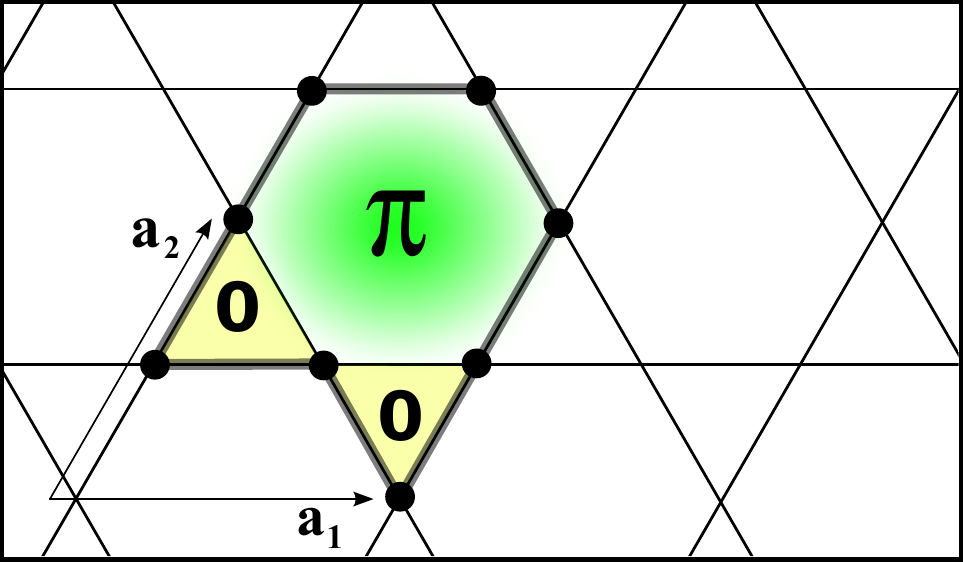}
\caption{\label{fig:kagome}
(Color online) The geometrical unit cell of the kagome lattice is shown as a shaded region,
along with the lattice vectors ${\bf a}_{1}$ and ${\bf a}_{2}$. The mean-field {\it Ansatz} 
of the $U(1)$ Dirac spin liquid has zero flux piercing the triangles and $\pi$ flux piercing 
the hexagons. Hence, each geometrical unit cell encloses a flux $\pi$, whose implementation 
requires a $2\times1$ expansion of the geometrical unit cell.}
\end{figure}

\floatsetup[table]{capposition=top}
\begin{table*}[t]
\centering
\begin{tabular}{lllllllll}
 \hline \hline
       \multicolumn{1}{l}{Size}
    & \multicolumn{1}{c}{$0$-LS}
    & \multicolumn{1}{c}{$1$-LS} 
    & \multicolumn{1}{c}{$2$-LS}
    & \multicolumn{1}{c}{$0$-LS}
    & \multicolumn{1}{c}{$1$-LS}
    & \multicolumn{1}{c}{$2$-LS}
    & \multicolumn{1}{c}{Ground state}  
    & \multicolumn{1}{c}{$S=2$ state}  \\ \hline
       
\multirow{1}{*}{$48$} & $-0.4293510(4)$ & $-0.4352562(3)$ & $-0.436712(1)$ & $-0.417980(1)$ & $-0.425218(1)$ & $-0.427155(3)$ & $\bm{-0.437845(4)}$ & $\bm{-0.43070(1)}$   \\ 
                                                                                                                       
\multirow{1}{*}{$108$} & $-0.4287665(4)$ & $-0.4341032(5)$ & $-0.435787(3)$ & $-0.425567(1)$ & $-0.431290(1)$ & $-0.433217(3)$ & $\bm{-0.437178(9)}$ & $\bm{-0.43516(2)}$   \\ 
                                                                                                                                          
\multirow{1}{*}{$192$} & $-0.4286749(4)$ & $-0.4334481(5)$ & $-0.435255(4)$ & $-0.427360(2)$ & $-0.432274(1)$ & $-0.434181(5)$ & $\bm{-0.43674(3)}$ & $\bm{-0.43597(3)}$   \\

\multirow{1}{*}{$432$} & $-0.428656(1)$ & $-0.432519(1)$ & $$ & $-0.428274(3)$ & $-0.432169(1)$ & $$ & $\bm{-0.43652(4)}$ & $\bm{-0.43631(4)}$   \\ \hline \hline

\end{tabular}
\caption{Energies of the $U(1)$ Dirac spin liquid (columns $2$-$4$) and its $S=2$ excited 
state (columns $5$-$7$), with $p=0$, $1$, and $2$ Lanczos steps on different cluster sizes 
obtained by variational Monte Carlo are given. The ground-state and $S=2$ excited-state 
energies of the spin-$1/2$ Heisenberg model estimated by using zero-variance extrapolation of 
variational energies on different cluster sizes are marked in bold.}
\label{tab:en-lanczos}
\end{table*}

For a precise identification of the spin liquid, the first step is to address the key 
issue of whether the ground state has a finite spin gap or not. Recent large scale DMRG 
calculations claim for a finite gap to spin excitations in the thermodynamic 
limit~\cite{Yan-2011,Depenbrock-2012,Nishimoto-2013,Jiang-2008}. However, the estimate of 
the triplet spin gaps given by these DMRG studies is not fully consistent. While 
conventional DMRG calculations performed on cylindrical geometries gave an estimate of 
$\Delta=0.13(1)$~\cite{Yan-2011,Depenbrock-2012}, a different estimate of $\Delta=0.055(5)$
was obtained by considering ``square'' clusters with periodic boundaries and preserving all
pointlike symmetries~\cite{Jiang-2008}. Moreover, more recent grand canonical DMRG calculations 
on an isotropic ``hexagonal'' cluster gave an estimate of $\Delta=0.05(2)$, which emphasized the 
importance of considering isotropic/symmetric clusters instead of cylindrical geometries, and 
of performing a simultaneous size scaling in all dimensions~\cite{Nishimoto-2013}. The message 
of these latter results is that the spin gap (if any) may be much smaller than what has been 
claimed by the standard DMRG. 

An alternative point of view is given by Gutzwiller projected fermionic wave functions that 
strongly support a gapless scenario described by an algebraic $U(1)$ Dirac spin liquid.
Indeed, explicit numerical calculations have shown the $U(1)$ Dirac spin liquid to be 
stable (locally and globally) with respect to dimerizing into all known valence-bond 
crystal phases~\cite{Ran-2007,Ma-2008,Iqbal-2011a,Iqbal-2012}. In addition, it was shown 
that, within this class of Gutzwiller projected wave functions, all the fully symmetric, 
gapped $\mathbb{Z}_{2}$ spin liquids have a higher energy compared to the $U(1)$ Dirac 
spin liquid~\cite{Iqbal-2011b,Lu-2011,Tay-2011,Yang-2012}. Only a minor energy gain can be obtained 
by fully relaxing all the variational freedom of the Gutzwiller projected wave function
(furthermore, this energy gain {\it decreases} upon increasing the cluster size)~\cite{Clark-2013}. 
Most importantly, it was shown that upon application of a couple of Lanczos steps on the 
$U(1)$ Dirac spin liquid, very competitive energies can be achieved, without 
disturbing the power-law decay of the long distance real space spin-spin correlations, i.e., 
retaining a gapless state~\cite{Iqbal-2013}. However, so far, a direct calculation of the spin 
gap has not been afforded within this approach.

In this Rapid Communication, we compute the $S=2$ spin gap $\Delta_2$ of the Heisenberg
model on the kagome lattice by considering spinon excitations around the Dirac nodes. 
We show that the simple variational wave function is gapless, implying that the Gutzwiller
projector does not affect the mean-field expectation. Most importantly, by applying a few 
Lanczos steps onto the variational states with $S=0$ and $S=2$, we can reach a reliable
estimation of $\Delta_2$ on several cluster sizes, so to extrapolate in the thermodynamic limit.
Our results are compatible with a gapless $S=2$ excitation, the upper bound of the gap being
$\Delta_2 \simeq 0.02$ (leading to a $S=1$ gap of $\Delta \simeq 0.01$, much smaller than 
what has been obtained by standard DMRG on cylindrical geometries~\cite{Yan-2011,Depenbrock-2012} 
and closer to other DMRG calculations performed on square
clusters~\cite{Nishimoto-2013,Jiang-2008}).

On the experimental front, recent neutron scattering measurements on single-crystal samples of 
the near perfect kagome spin-$1/2$ Heisenberg model compound ${\rm ZnCu_{3}(OH)_{6}Cl_{2}}$ point 
towards a gapless spin-liquid behavior, with an upper bound of the spin gap which is estimated to 
be $\sim J/10$ (if a gap exists)~\cite{Han-2012}.

\begin{figure}[b]
\includegraphics[width=1.0\columnwidth]{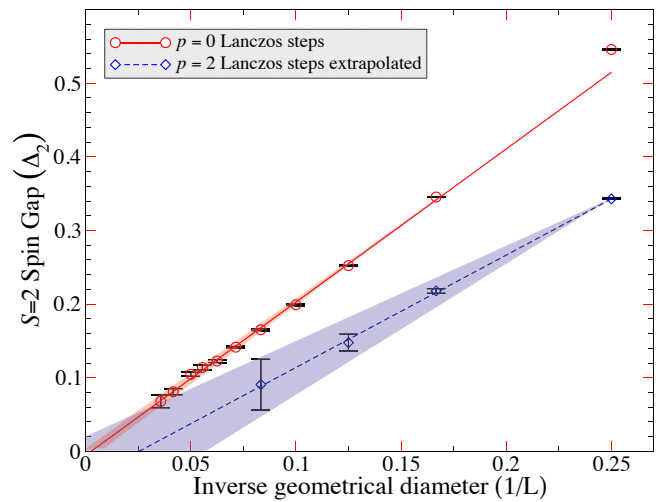}
\caption{\label{fig:Gap-FSS}
(Color online) Finite-size scaling of the $S=2$ spin gap as a function of the inverse geometrical 
diameter ($1/L$). Both the $p=0$ and $p=2$ extrapolated values give an estimate which is zero 
(within error bars) in the thermodynamic limit. A linear fit is used in both cases. The largest 
size considered for the $p=0$ case is $2352$ sites.}
\end{figure}

\begin{figure*}[t]
\includegraphics[width=0.49\columnwidth]{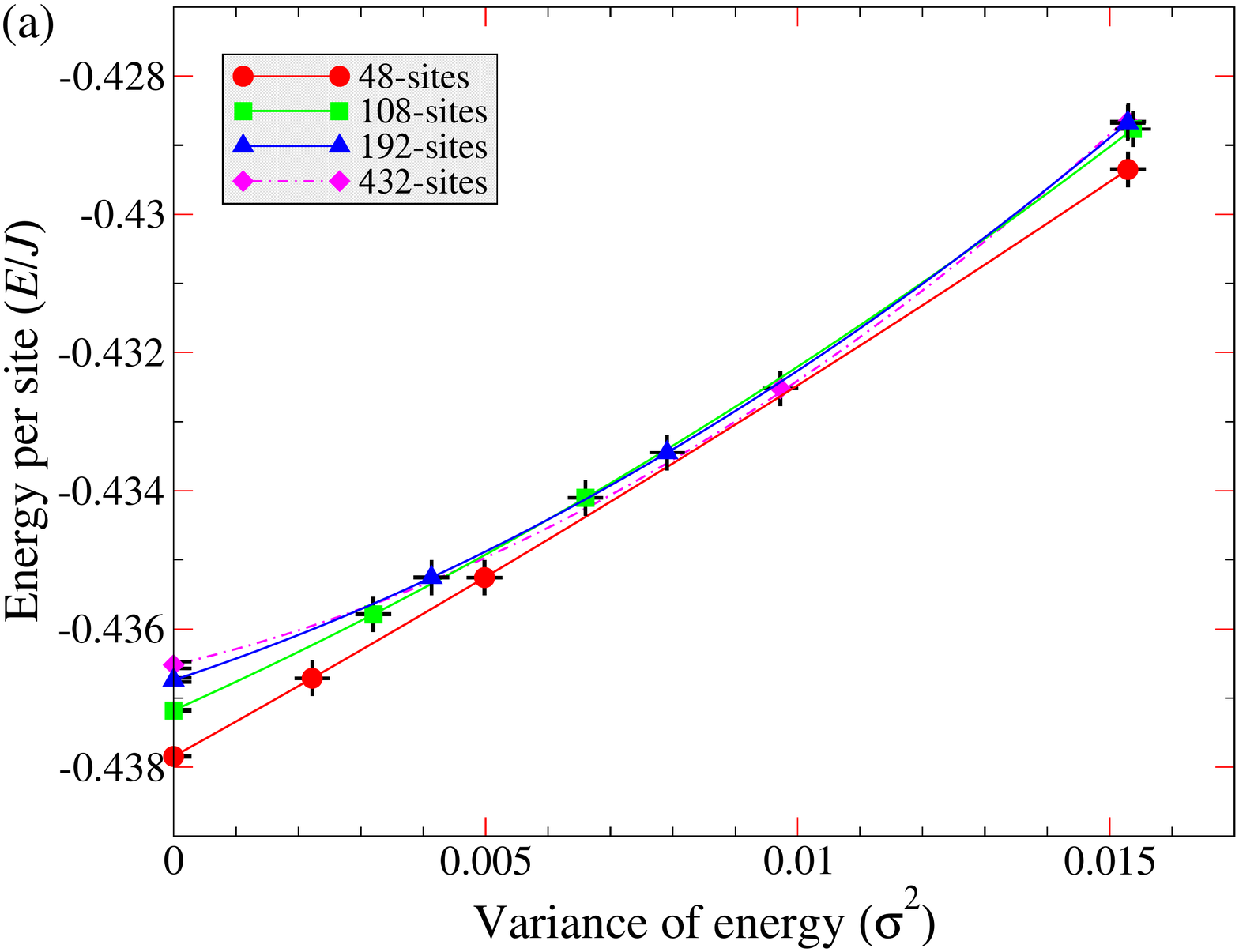}\label{fig:GS-ZVE}\quad
\includegraphics[width=0.49\columnwidth]{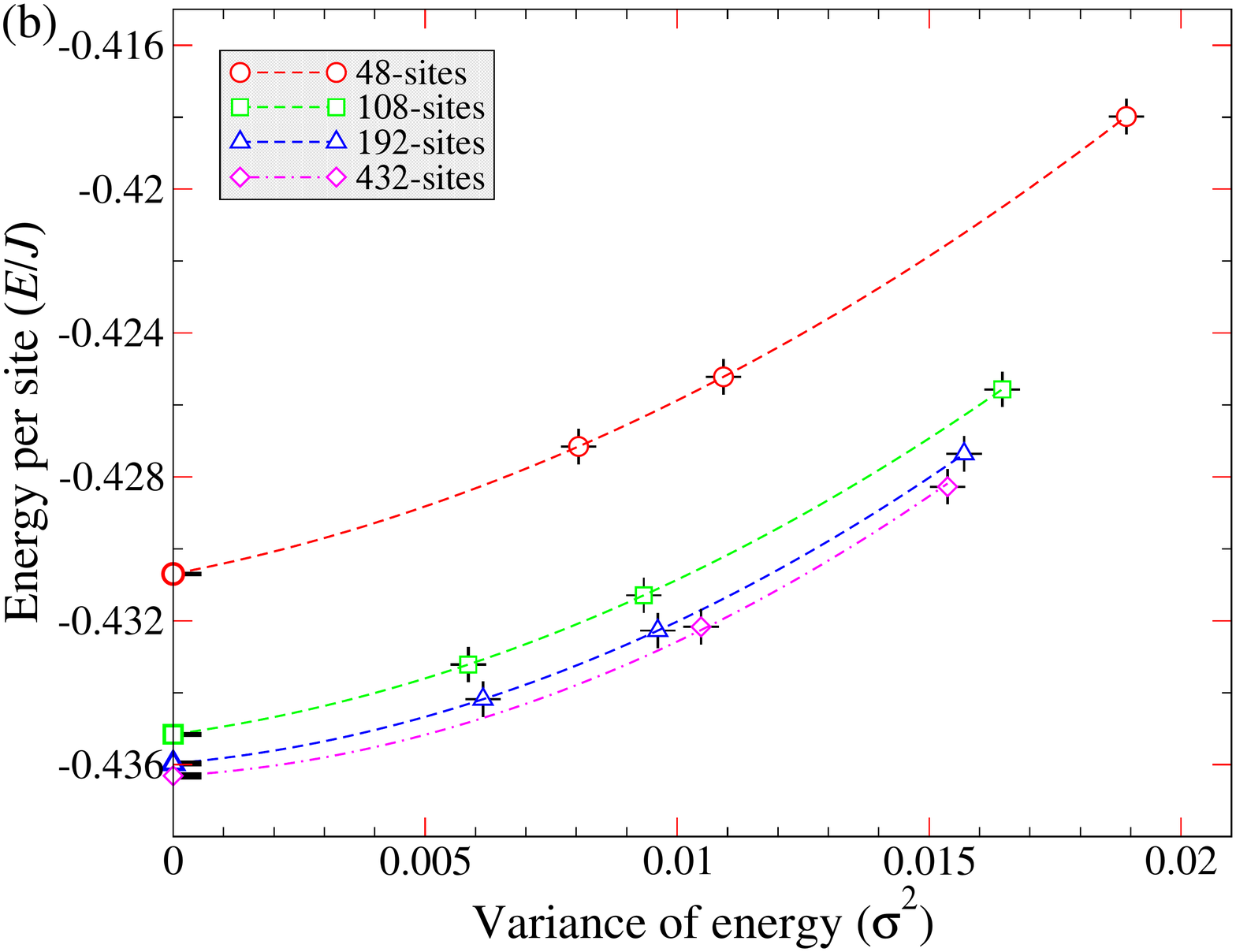}\label{fig:S2-ZVE}
\caption{\label{fig:ZVE}
(Color online) (a) Variational energies as a function of the variance of energy, for zero, one, 
and two Lanczos steps. The ground-state $S=0$ energy on the $48$-, $108$-, and $192$-site clusters 
is estimated by extrapolating the three variational results to the zero-variance limit by a quadratic 
fit, while only two points have been used for the $432$-site cluster, and the method of extrapolation 
is explained in the main text. (b) The same for the $S=2$ excited state.}
\end{figure*}

{\it Model, wave function, and numerical technique}.
The Hamiltonian for the spin-$1/2$ quantum Heisenberg antiferromagnetic model is
\begin{equation}
\label{eqn:heis-ham}
\hat{{\cal H}} = J \sum_{\langle ij \rangle} \mathbf{\hat{S}}_{i} \cdot \mathbf{\hat{S}}_{j},
\end{equation} 
where $J>0$ and $\langle ij \rangle$  denotes the sum over nearest-neighbor pairs of sites. 
The $\mathbf{\hat{S}}_{i}$ are spin-$1/2$ operators at each site $i$. All energies will be 
given in units of $J$.

The variational wave function is constructed by projecting a noncorrelated fermionic state:
\begin{equation}
\label{eqn:var-wf}
|\Psi_{{\rm Dirac}}\rangle=\mathbf{{\cal P}_{G}}|\Psi_{{\rm MF}}\rangle,
\end{equation}
where $\mathbf{{\cal P}_{G}}=\prod_{i}(1-n_{i,\uparrow}n_{i,\downarrow})$ is the full 
Gutzwiller projector enforcing the one fermion per site constraint. Here, 
$|\Psi_{{\rm MF}}\rangle$ is the ground state of a mean-field fermionic Hamiltonian
that contains hopping only:
\begin{equation}
\label{eqn:MF0}
\hat{{\cal H}}_{{\rm MF}} = \sum_{i,j,\alpha} \chi_{ij} \hat{c}_{i,\alpha}^{\dagger}\hat{c}_{j,\alpha},
\end{equation}
where $\alpha=\uparrow,\downarrow$ and $\chi_{ij}=\chi_{ji}$ (i.e., real hopping). 
The $U(1)$ Dirac state is defined with nontrivial gauge magnetic fluxes piercing the triangles 
and hexagons (see Fig.~\ref{fig:kagome}~\cite{Ran-2007}) We would like to emphasize that the 
projected wave function does not contain any variational parameter and it is fully determined by
fixing the flux pattern. The $S=0$ {\it Ansatz} is constructed by filling the lowest-energy
single-particle states for both up and down electrons; suitable boundary conditions must
be considered in order to have a unique mean-field state. This simple {\it Ansatz} for the 
ground state of the kagome spin-$1/2$ Heisenberg model gives rather good accuracy on the energy 
per site in the thermodynamic limit, i.e., $E/J=-0.428~67(1)$ [to be compared with $E/J=-0.4386(5)$
obtained by the DMRG approach~\cite{Depenbrock-2012} and $E/J=-0.4364(1)$ by PESS~\cite{Xie-2013}]. 
The $S=2$ state is constructed by changing boundary conditions, in order to have four spinons 
in an eightfold degenerate single-particle level at the chemical potential; a unique mean-field 
state is then obtained taking all these spinons with the same spin (so that the total wave 
function has $S=2$). The $S=2$ state has $k=(0,0)$ and is particularly simple since it can be
represented with a single Slater determinant (as the ground state).

In order to have a systematic improvement of the trial variational wave function and 
approach the true ground state, we can apply a few Lanczos steps to 
$|\Psi_{\rm Dirac}\rangle$~\cite{Sorella-2001}:
\begin{equation}
\label{eqn:psi-ls}
|\Psi_{p\text{-}\rm{LS}}\rangle =  \bigg{(}1+\sum_{k=1}^{p}\alpha_{k}\hat{{\cal H}}^{k}\bigg{)}|\Psi_{\rm Dirac}\rangle.
\end{equation}
Here the $\alpha_{k}$'s are variational parameters. The convergence of 
$|\Psi_{p\text{-}\rm{LS}}\rangle$ to the exact ground state $|\Psi_{\rm ex}\rangle$ 
is guaranteed for large $p$ provided the starting state is not orthogonal to
$|\Psi_{\rm ex}\rangle$, i.e., for $\langle\Psi_{\rm ex}|\Psi_{\rm Dirac}\rangle \neq 0$. 
On large cluster sizes, only a few steps can be efficiently performed and here we implement 
the case with $p=1$ and $p=2$ ($p=0$ corresponds to the original trial wave function). 
Subsequently, an estimate of the exact ground-state energy may be achieved by the method of 
variance extrapolation: For sufficiently accurate states, we have that 
$E\approx E_{\rm ex}+{\rm constant}\times\sigma^{2}$, where 
$E=\langle\hat{{\cal H}}\rangle/N$ and 
$\sigma^{2}=(\langle\hat{{\cal H}}^{2}\rangle-\langle\hat{{\cal H}}\rangle^{2})/N$ 
are the energy and variance per site, respectively, whence, the exact 
ground-state energy $E_{\rm ex}$ can be extracted by fitting $E$ vs 
$\sigma^{2}$ for $p=0$, $1$, and $2$. The energy and its variance for $p=0$, $1$, and $2$ 
Lanczos steps are obtained using the standard variational Monte Carlo method, independently 
for $S=0$ and $S=2$ states. 

\floatsetup[table]{capposition=bottom}
\begin{table}[b]
\centering
\begin{tabular}{lcccc}
 \hline \hline
       \multicolumn{1}{l}{$48$}
    & \multicolumn{1}{c}{$108$}
    & \multicolumn{1}{c}{$192$} 
    & \multicolumn{1}{c}{$432$} 
    & \multicolumn{1}{c}{$\infty~2D$ limit} \\ \hline
       
\multirow{1}{*}{$0.3429(7)$} & $0.218(3)$ & $0.147(11)$ & $0.090(34)$ & $-0.04(6)$ \\ \hline \hline

\end{tabular}
\caption{The $S=2$ gap of the spin-$1/2$ Heisenberg model obtained from the estimates of $S=0$ 
and $S=2$ energies on different cluster sizes (marked in bold in Table~\ref{tab:en-lanczos}) 
is given. The estimate in the thermodynamic limit is zero (within error bars).}
\label{tab:Gap}
\end{table}

{\it Results}.
Our variational calculations were performed on square (i.e., $3\times L \times L$) 
clusters with periodic boundaries. We start by computing the $S=2$ gap for the Dirac wave function 
$|\Psi_{{\rm Dirac}}\rangle$. Before Gutzwiller projection, the mean-field state 
$|\Psi_{{\rm MF}}\rangle$ is gapless; however, given the $U(1)$ low-energy gauge structure of 
the mean-field {\it Ansatz}, it is not obvious that this property is preserved after 
projection~\cite{Wen-2002}. In fact, the $U(1)$ gauge fluctuations are expected to be wild, 
possibly leading to some instability of the mean-field {\it Ansatz}~\cite{Hermele-2004};
nevertheless, it has been shown that $|\Psi_{{\rm Dirac}}\rangle$ is essentially stable 
against all possible perturbations (only a marginal improvement has been obtained by a
totally unconstrained optimization of the pairing function~\cite{Clark-2013}). Here, we
show that the $S=2$ gap is vanishing in the thermodymanic limit (see Fig.~\ref{fig:Gap-FSS}).
This result is interesting in itself, since it clearly shows that the Gutzwiller projection
does not alter the low-energy properties of the mean-field state; this outcome is compatible 
with the fact that the spin-spin correlations have a power-law decay at long distances, namely, 
$\langle \mathbf{\hat{S}}_{r} \cdot \mathbf{\hat{S}}_{0}\rangle \sim 1/r^4$~\cite{Hermele-2008}.  

Let us move to the central part of this work by applying a few Lanczos steps to the original
Dirac state. In Table~\ref{tab:en-lanczos} and Fig.~\ref{fig:ZVE}, we report the results for 
the $S=0$ and $S=2$ states separately. The very computationally demanding $p=2$ calculations 
have been performed for $48$, $108$, and $192$ sites, while for the $432$-site cluster only the 
first Lanczos step has been considered. In the former cases, the zero-variance extrapolation 
can be exploited by a quadratic fit of the three points, namely, 
$E=E_{\rm ex}+{\mathcal A}\times\sigma^2+{\mathcal B}\times(\sigma^2)^2$.
The zero-variance extrapolation gives size consistent results for the energy 
per site~\cite{Hu-2013}. Indeed, even though the Lanczos step procedure (with a fixed $p$) 
becomes less and less efficient when increasing the system size, the extrapolation procedure 
remains accurate: This can be seen by noticing that the gain in the energy and variance with 
respect to $p=0$ decreases with $L$, but the extrapolation is not affected, since the slope
is essentially unchanged. For the larger cluster, i.e., $432$ sites, we also considered a 
quadratic fit: We first obtained an estimate of the ${\mathcal A}$ and ${\mathcal B}$ coefficients 
by a size scaling of the smaller clusters and then verified that indeed, these values give an 
excellent fit (i.e., least mean-square error) of the points for $432$-site cluster. 

The computation of the $S=0$ and $S=2$ energies allows us to obtain the {\it extrapolated} gap 
for each size independently, which is reported in Table~\ref{tab:Gap} and Fig.~\ref{fig:Gap-FSS}. 
Here, despite having an error bar that increases with the system size, we can reach an extremely 
accurate thermodynamic extrapolation, namely, $\Delta_2=-0.04 \pm 0.06$. Therefore, our main 
conclusion is that our competitive {\it Ansatz} has gapless excitations. 
Our best estimate for the upper bound on the $S=2$ gap is $\Delta_2 \simeq 0.02$, leading to a 
$S=1$ gap that would be approximately half of this value, i.e., $\Delta \simeq 0.01$. This latter 
result is much lower than previous DMRG estimates~\cite{Yan-2011,Depenbrock-2012}, which were 
done by considering cylindrical geometries $L_x \times L_y$, i.e., first performing the limit 
$L_x \to \infty$ and then increasing the circumference $L_y$. In contrast to this, it has been 
shown that more isotropic lattices lead to a different thermodynamic extrapolation, suggesting
possible difficulties when using large aspect ratios $L_x \gg L_y$, and nonsimultaneous size 
scaling of different dimensions~\cite{Nishimoto-2013}.
This fact has also been discussed in a recent work on valence-bond solids~\cite{Sandvik-2012},
where it has been pointed out that the long cylinders may turn even a true valence-bond
order into a disordered phase (with only short-range correlations). 

{\it Summary}.
We have reported the $S=2$ gap in the kagome spin-$1/2$ Heisenberg antiferromagnetic model by 
using an improved variational technique based upon Gutzwiller projected fermionic wave functions. 
The application of a few Lanczos steps on top of the $U(1)$ Dirac spin liquid, together with a 
zero-variance extrapolation, gives extremely accurate results, which strongly support the fact 
that the ground state is gapless. 
However, controversial claims based on DMRG~\cite{Yan-2011,Depenbrock-2012,Ju-2013}
and tensor network methods~\cite{Poilblanc-2012,Xie-2013} of a gapped topological spin
liquid raise concerns about the possibility for the existence of different basins of
attraction in the energy landscape. In this picture, it would be very difficult to cross
over from one state to the other, requiring either a very large number of states within
DMRG and tensor network methods or a very large number of Lanczos steps in our approach.

{\it Acknowledgments}.
We acknowledge P.~A.~Lee for stimulating discussions during the KITP program ``Frustrated 
Magnetism and Quantum Spin Liquids: From Theory and Models to Experiments'' and S.~Sorella for 
several suggestions during the accomplishment of this work. This research was supported in part 
by the National Science Foundation under Grant No. NSF PHY11-25915 and by PRIN 2010-11.
Partial support by the ``Agence Nationale de la Recherche'' under Grant
No. ANR 2010 BLANC 0406-0 and by the CALMIP supercomputer
center (Toulouse, France) are also acknowledged.

\newpage

\end{document}